\documentclass[fleqn,twoside]{article}
\usepackage{espcrc2}
\usepackage{amsmath}
\usepackage{graphicx}
\usepackage[figuresright]{rotating}

\def\prl{{\it Phys. Rev. Letters\ }}

\title{Gamma rays precursors  and afterglows  surrounding UHECR events: Z-burst model is still alive }
\author{D.Fargion\address[MCSD]{Physics Department Rome Univeristy\\
INFN,Rome, Italy} and A. Colaiuda}

\begin{document}

\begin{abstract}

The Z-burst model and the direct propagation of UHE proton in
negligible extragalactic magnetic fields produce gamma-rays
afterglows and precursors halos, respectively at GeVs and TeV
energy band a few degree around the UHECR arrival direction. The
possible correlation of UHECR clusters (doublet, triplet) with
nearby BL Lac sources at $E_p \simeq 4 \cdot 10 ^{19}$ eV offer a
test for this necessary Gamma-UHECR trace. We estimate the
secondary gamma energy and spectra and we suggest how to
disentangle between the different scenarios. We show why Z-Burst
model is still the  most realistic model to explain UHECR
behaviour and their correlation to known BL Lac sources.
\vspace{1pc}
\end{abstract}

\maketitle

\section{Introduction}
The UHECR events with energy above $4\times10^{19}$ eV are bounded
by the primordial photon drag ( the well known GZK cut-off) in a
very narrow universe ($\sim 50 Mpc$). Because of their charge
UHECR are bent and blurred by cosmic magnetic fields. However,
UHECR because of their extremely rigidity maintain  their
primordial arrival direction. Surprisingly, there aren't any
nearby known galactic structures or super-galactic objects
correlated with these UHE events which can accelerate protons at
so high energy. Recent AGASA evidence of UHECR events clustering
toward some BL Lacs (see table \ref{sorg}) seems to confirm a
cosmic extra-galactic origin of UHECR from compact sources. So
there is a puzzle to solve: if these UHECR (just at the edge of
GZK cut off energies) are protons ( as experimental data suggest),
how can they  point to their primary BL Lac sources (often at
large distance) with a small ($<2.5^o$) angular dispersions? There
are two mechanisms to solve this puzzle: the rectilinear
propagation of UHE proton from the source to the Earth (proposed
by Berezinsky) within negligible extragalactic magnetic fields and
the Z-burst model \cite{Fargion-Mele-Salis99},\cite{Weiler99}
(based on UHE ZeV neutrino scattering on relic light neutrinos in
hot dark halos). Both the processes mainly lead ,as a by product,
to a signal respectively in TeV (direct flight) and GeV  (Z-Burst)
photons. These photons can be observed either as afterglows or as
precursors of the UHECR events. They may be (respectively) long
lived showers (direct flight) of a few years or months or prompt
signals (Z-Burst) within days or hours. In the following, we will
study these two different models and we will show that direct
proton flight call for unobserved TeV tails while  the Z-burst
model is still the most competitive and realistic process to solve
the GZK puzzle because it agrees with the observed GeV gamma tails
along the BL Lac Sources correlated to UHECR cluster events.
\begin{table*}[htb]
\caption{BL Lacs correlated with UHECR. We  focus and remind among
the ($14$) clusters here only those ($9$) BL Lacs
   whose distance is above z=0.1 and whose energy is well defined}\label{sorg}
\newcommand{\m}{\hphantom{$-$}}
\newcommand{\cc}[1]{\multicolumn{1}{c}{#1}}
\renewcommand{\tabcolsep}{2pc} 
\renewcommand{\arraystretch}{1.2} 
\begin{tabular}{@{}lllll}
\hline
EGRET name & Possible BL Lac & l($^ \circ$) & b ($^ \circ$)& Redshift z \\
    \hline
    0433+2908 & 2EG J0432+2910* & 170.5 & -12.6 & - \\
    0808+5114 & 1ES 0806+524* & 166.2 & 32.91 & 0.138 \\
    1009+4855 & GB 1011+496 & 165.5 & 52.71 & 0.2 \\
    1222+2841 & ON 231* & 201.7 & 83.29 & 0.102 \\
    1424+3734 & TEX 1428+370 & 63.95 & 66.92 & 0.564 \\
    1605+1553 & PKS 1604+159* & 29.38 & 43.41 & - \\
    1850+5903 & RGB J1841+591 & 88.68 & 24.29 & 0.53 \\
    2352+3752 & TEX 2348+360 & 109.5 & -24.91 & 0.317 \\
    1052+5718 & RGB J1058+564* & 149.6 & 54.42 & 0.144 \\
    \hline
    \end{tabular}\\[2pt]
    \end{table*}

 \section{The rectilinear propagation of UHECR}

The rectilinear propagation of UHECR protons is possible only if
the intergalactic magnetic field is extremely weak. The actual
knowledge of the intergalactic magnetic field is still very poor.

Usually, extragalactic magnetic fields are characterized by an
average strength B and by a coherence length $l_c$ which means
that its power spectrum has a cut-off at the wavelength space
$k=2\pi/l_c$.

Neglecting the energy losses for the moment, the r.m.s. deflection
angle over a distance d in such field is \cite{bh00}:

\begin{equation}\label{eqn:deflessione1} \begin{split}
&\vartheta(E,\mathit{d})\simeq\frac{(2\mathit{dl_c}/9)^{1/2}}{R_L}\\
&\simeq0.8^\circ(E_{20})^{-1}(d_{10})^{1/2}(l_{c1})^{1/2}(B_{\bot-9})
 \end{split}\end{equation}
 where $E_{20}$ is the energy of the proton, $E_p=10^{20}$ eV,
 $d_{10}$ is the distance, $d=10$ Mpc and $l_{c1}$ is the coherence
 length of magnetic field, $l_{c1}=1$ Mpc.
  It should be noticed that a Galactic Magnetic field  $B_G \simeq \mu
  Gauss$ and a coherent lenght of $l_c \simeq 100 pc$, within a galactic size of $d\simeq 10kpc$ already
  implies an angular rms of the order $\vartheta(E,\mathit{d}) \simeq 2.4^o$.
 The deflection implies also an average time delay, $\tau$,
 relative to  rectilinear propagation:
 \begin{equation} \label{eqn:ritardo} \begin{split}
 &\tau(E, d)\simeq\frac{d\vartheta(E,d)^2}{4c} \\
&\simeq1.5\times10^{3}\biggl(\frac{E}{10^{19}eV}\biggr)^{-2}\biggl(\frac{d}{1000Mpc}\biggr)^2\cdot
\\&\cdot\biggl(\frac{l_c}{10Kpc}\biggr)\biggl(\frac{B_\bot}{10^{-11}G}\biggr)^2yr
\end{split}\end{equation}
The Galactic Magnetic field  $B_G \simeq
\mu
  Gauss$ and as before a coherent lenght of $l_c \simeq 100 pc$,  a galactic size of $d\simeq 10kpc$ already
  implies a time dilution of the order $\tau(E, d) \simeq 15 yrs $.
Berezinsky required \cite{Berezinsky 2002}, for having rectilinear
propagation of UHECR with an energy $E\sim 10^{19}$eV from a
distance  $d\sim 1000Mpc$, that the angular deflection produced by
the magnetic field is not larger than the angular resolution of
sources in the detectors ( typically $\vartheta_{res}=2.5^\circ$).
According to equation \ref{eqn:deflessione1}, this request gives
an upper limit to the strength of a magnetic field with a small
coherence length ( $l_c=10Kpc$):
\begin{equation}\label{campomagnetico}
B\leq 3\times10^{-10}G
\end{equation}

For a proton with an energy $E_p=4\times10^{19}$eV the relevant
mechanism of energy losses is the photo electron pair production
with the photons of the Cosmic Microwave Background (CMB). In the
following we refer to the electron and to the positron as
electron. The electron pairs are produced by  a virtual photon of
characteristic energy:
\begin{equation}\label{eqn:compton}\begin{split}
   \hbar\varpi=&\frac{4}{3}\gamma^2(\hbar\omega)_{CMB}\\&\sim1.35\cdot10^{18}(\hbar\omega)_{CMB}\biggl(\frac{E_p}{4\cdot10^{19}eV}\biggr)^2eV
\end{split} \end{equation} where $\gamma$ is the Lorentz factor of
protons and $\hbar\omega_{CMB}$ is the average energy of
$\gamma_{CMB}$ that is $6.34\times10^{-4}$eV. The electron will
have half of the energy of this photon i.e.:
\begin{equation}\label{eqn:mezzo}
E_{e_1}=\frac{1}{2}\hbar\varpi\sim6.8\times10^{17}eV\biggl(\frac{E_p}{4\cdot10^{19}eV}\biggr)^2
\end{equation}
At this energy and in the center of moment (mass) frame system,
the electron will behave nearly as a massless particle because the
quantity
\begin{equation}\label{eqn:centromassa}\begin{split}
&E_{cms}=\\&=11.2\biggl[(\hbar\omega)_{CMB}\biggl(\frac{E_e}{6.8
\times 10^{17}}\biggr)\biggr]^{1/2}TeV
\end{split}\end{equation}
is $\gg m_e$. So the electron will distribute half of his energy
to a new electron ($e_2$) and half to a new photon ($\gamma_2$).
The $\gamma_2$ will produce a new pair of $e^+e^-$ while the $e_2$
will produce a new $\gamma_3$ and a new $e_3$.

This process will form an electromagnetic shower by I.C.S.-pair
production  and (in analogy to the Heitler for normal showers in
atmosphere)
 it will continue since $E_{cdm}\sim m_e$;  it will
happen for a transition energy around $E_e=7.5\times10^{13}$ eV.

\begin{figure}[htb] \vspace{9pt}

\includegraphics[width=70mm]{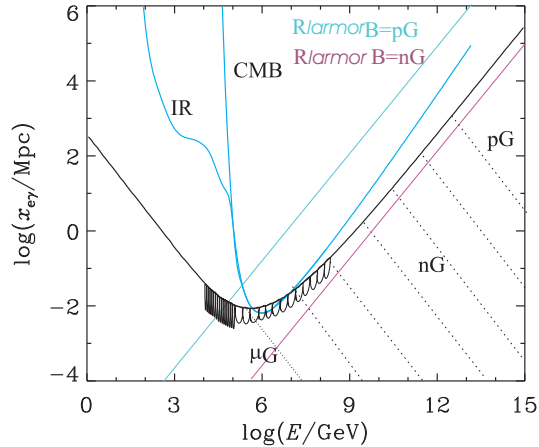}

\caption{Qualitative representation of the halving's and ICS
processes for electrons; we also trace the Larmor radius for
$B=10^{-12}G$ and $B=10^{-9}G$ and by dotted lines the energy
losses for synchrotron radiation} \label{elettroni}
\end{figure}

\begin{figure}[htb] \vspace{9pt}

\includegraphics[width=70mm]{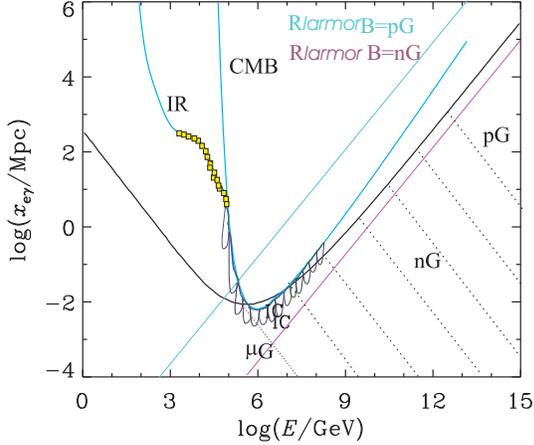}

\caption{Qualitative representation of the energy's halving and
ICS processes for photons; we also trace by different colors  the
Larmor radius for $B=10^{-12}G$ and $B=10^{-9}G$ and by dotted
lines the energy losses for synchrotron radiation} \label{fotoni}
\end{figure}

For this energy, we have an effective Inverse Compton scattering
that will produce only photons with an energy  lower and lower
than the incoming electron:
\begin{equation}\label{eqn:comptonelettrone}\begin{split}
&\hbar\varpi=\frac{4}{3}\gamma^2(\hbar\omega)_{CMB}\\&\sim3.23\cdot10^{11}(\hbar\omega)_{CMB}\biggl(\frac{E_e}{10^{13}eV}\biggr)^2eV
\end{split}\end{equation}
The energy fraction given by electron to the photon reduces itself
as the process goes on: the final result is the production (and
the consequent pilling up) of TeV photons; the electron pairs
progenitors aren't much deviated by the necessary small
extragalactic magnetic field: in fact, the electrons are very
collimated with the proton directions until $ E_e=10^{15}eV$
because their Larmor's radius is much smaller than their energy
losses (by I.C.S.) length. After this energy, their Larmor's
radius is:
\begin{equation}\label{eqn:larmor0}
R_L=20\biggl(\frac{B}{10^{-11}G}\biggr)^{-1}\biggl(\frac{E}{10^{14}eV}\biggr)Kpc
\end{equation}
and so their deviation is  still quite negligible respect to their
short life time distance by ICS with the CMB $l_e\sim c\tau_e\sim
5(E/10^{14}eV)^{-1}Kpc$. This can be seen in figure
\ref{elettroni} and in figure \ref{fotoni} that describe the
process of halving and the ICS respectively for electrons and
photons.
 The TeV photons might suffer of an additional and
partial (for hundreds or Mpc distances) IR cut-off; but mostly
these TeV traces might survive and arrive  for a long times
(years) collinear with the UHECR event. This TeV-UHECR connection,
(except for a questionable mild TeV clustering around $TEX
1428+370 $ ) is not observed in known (Milagro \cite{atkin}) map.

\section{The Z-burst model}

\begin{figure}[htb] \vspace{9pt}

\includegraphics[width=70mm]{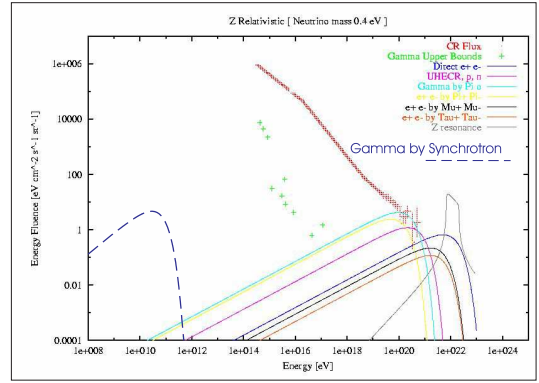}

\caption{Zeta resonance and main secondaries for $m_\nu = 0.4$}
\label{zeta}
\end{figure}

A different solution of the UHECR puzzle above the GZK cut-off,
able to explain also the BL Lacs correlation with UHECR, is based
on light relic neutrino masses in hot halo, a wide calorimeter for
UHE neutrinos messengers.  The interaction between an
extragalactic ZeV UHE neutrino and a halo of relics neutrinos with
a fixed mass may produce Z-resonance whose boosted decay  in
flight leads to  the observable UHECR nucleons (proton,
anti-protons, neutron, anti-neutrons). In the s-channel the
interaction of neutrinos of the same flavor occurs by Z-exchange:
$ \nu\bar{\nu_r}\rightarrow Z^*$ (see figure \ref{zeta}).
The cross section of this reaction shows a peak due to the
resonant Z-production at $s=M^2_Z$: this Z-resonance ( here
averaged on the energy) requires an initial neutrinos energy
$E_\nu\sim 10^{22}(m_\nu/0.4eV)^{-1}$eV  to produce a proton with
energy $E_p=2.2\times10^{20}(m_\nu/0.4eV)^{-1}eV$ ( see table
\ref{molteplicita}) \cite{Fargion03}. It's important to evidence
that a $ m_\nu \sim 0.4eV$ gives a relic neutrino number density:
$$n_{\nu_r} \sim 4.725\times10^{2}\biggl(\frac{m_\nu}{0.4eV}\biggr)^3\biggl(\frac{V}{300km/s}\biggr)^3cm^{-3}$$
for halo's size $l\sim 30Mpc$, compatible with a Local Super
Cluster. So we have a number density contrast $\sim 8$ above
cosmic background ($n_\nu^0\sim 56cm^{-3}$). A similar contrast is
also available for a mean neutrino velocity $2$ times larger
(corresponding to our Sun motion in the Black Body Radiation) and
a relic neutrino mass as small as half ($m_\nu \simeq 0.2$ eV) the
assumed one. This light mass is better comparable with observed
atmospheric neutrino's mass splitting ($\Delta_m\sim 0.05$ eV).
Non degenerated lightest neutrino masses ($m_\nu\leq 0.1$ eV)
would lead to multi-bump uhecr modulation at highest energy
(\cite{Fargion-Mele-Salis99},\cite{Fargion2001})
\begin{table*}[h!]
\caption{Secondaries produced in the interaction $\nu\bar{\nu}
\rightarrow Z$, assuming $E_\nu=10^{22}$ eV, an incoming energy
neutrino fluence $F_\nu=2000 eVcm^{-2}s^{-1}sr^{-1}$ and a relic
neutrino mass $m_\nu=0.4eV$; the neutrino interaction probability
corresponds to $1\%$ }\label{molteplicita}
\newcommand{\m}{\hphantom{$-$}}
\newcommand{\cc}[1]{\multicolumn{1}{c}{#1}}
\renewcommand{\tabcolsep}{1pc} 
\renewcommand{\arraystretch}{1.2} 
\begin{tabular}{@{}llllll}

  \hline
   & Multiplicity & Energy(\%) & $\sum E_{CM}$(GeV) & Peak energy (EeV) &$ \frac{dN}{dE}E^2(eV)$  \\
  \hline
  p & 2.7 & 6\% & 5.4 & $2.2\times10^{2}$& 1.2\\
  $\pi^0$ & 13 & 2\.4\% & 19.25 & 1.9$\times10^2 $& 4.25 \\
  $\gamma_{\pi^0}$ & 26 & 21.4\% & 19.25 & 95 & 4.25 \\
  $\pi^\pm$ & 26 & 42.8\% & 38.5 & 1.9$\times10^2$ & 4.25 \\
  ($e^+e^-)_{\pi}$ & 26 & 12\% & 11 & 50 & 2.3 \\
  ($e^+e^-)_{prompt}$ & 2 & 3.3\% & 2.7 & 10$\times10^3$ & 1.32 \\
  ($e^+e^-)_{\mu}$ & 2 & 1.1\% & 0.9 & 1.6$\times10^3$ & 0.45 \\
  ($e^+e^-)_{\tau}$ & 2 & 1.5\% & 1.3 & 1.2$\times10^3$ & 0.6 \\
  \hline
\end{tabular}
\end{table*}

A proton, with energy $E_p=2.2\times10^{20}eV$ at a distance of
30Mpc, suffers the $\gamma_{CMB}$ opacity and loses for photo-pair
production a factor ($e^{2.6} = 13.46 $) of its initial energy. So
there is a degradation of proton's energy around $E_{fp}\simeq
10^{19}eV$: this energy is compatible with the energy of UHECR
clustered events correlated with the observed BL Lacs  in table
\ref{tab:energiaalta}. At the Z-resonance we have also the
production of UHE electrons with energy $E_e=2\times10^{19}eV$ due
to charged pions decay. Assuming an extragalactic magnetic field
$B=10^{-9} G$ (according to experimental measure of Faraday
rotation and CMB anisotropy), the electrons interaction with it
will lead to $ E_\gamma=27.2 GeV$ photons direct peak:
\begin{equation}\label{eqn:gammasincro}\begin{split}
&E^{syn}_\gamma=\frac{3}{2}\gamma^2\biggl(
\frac{eB\hbar}{m_e}\biggr)\\&\sim
27.2\biggl(\frac{E_e}{2\times10^{19}eV}\biggr)^2\biggl(\frac{m_\nu}{0.4eV}\biggr)^{-2}\biggl(\frac{B}{nG}\biggr)GeV
\end{split}\end{equation} This mechanism is effective and fast and
it gives gamma-rays afterglows and precursors of UHECR events.  In
fact, we have an immediate emission of photons to tens of GeV as
soon as the electron interacts with the extragalactic magnetic
field and a subsequent emission of GeV photons until the electron
arrives at an energy$\sim 1.9\times10^{18}eV$. At those energies
the synchrotron radiation is leading to $E^{syn}_\gamma\simeq 270
$ MeV  emission. After this point, the dominant process will be
Inverse Compton scattering : ICS  leads to an electromagnetic
cascade to TeV energies  in analogy (but at a much lower level) to
what occurred for the UHE proton-electron pair showering in the
Berezinsky scenario.

In the Z-burst, there is a marginal (3.3\%)  production of direct
UHE electron pairs with $E\sim 5\cdot 10^{21}eV$ that create PeV
photons for synchrotron's radiation:
\begin{equation}\label{eqn:gammasincro3}
E^{syn}_\gamma=1.7 \biggl(\frac{E_e}{5 \cdot
10^{21}eV}\biggr)^2\biggl(\frac{m_\nu}{0.4eV}\biggr)^{-2}\biggl(\frac{B}{nG}\biggr)PeV
\end{equation}

 Nevertheless, this process is less significative and
probable than the process explained above for UHE electrons in the
s-channel.  Also these PeV photons are screened and degraded by
CMB cut off, to eV energy band. In fact, these photons will
produce $e^+e^-$ pairs for interaction with $\gamma_{CMB}$: the
electrons will have an energy $E_e\sim 0.85$ PeV and they will
produce via synchrotron radiation with the extragalatic magnetic
field $B=10^{-9}$ G photons with energy:
\begin{equation}\label{eqn:gammasincro4}
E^{syn}_\gamma=50B_{-9}\biggl(\frac{E_e}{0.85\cdot
10^{15}eV}\biggr)^2\biggl(\frac{m_\nu}{0.4eV}\biggr)^{-2}eV
\end{equation}
Nevertheless, we must remember that because of $\rho_{B_{-9}}\sim
10^{-7}\rho_{CMB}$  the energy losses are essentially given by ICS
as for the Berezinsky scenario and, in conclusion, we have the
sequent process  $E^{syn}_\gamma\rightarrow e^+e^- \rightarrow
E_{IC}$ that transforms the $e^+e^-$ energy to TeV photons with an
efficacy $\eta\sim 3/24\sim 12.5\%$ of the Z-boson's energy.

\section{Z-burst {\it vs.} UHE rectilinear propagation}

It's important to evidence some problems of the rectilinear
propagation of UHE :
\begin{enumerate}
    \item if we consider the  BL Lacs correlated with UHECR, we
    see that the sources, of which we know the energy and the
    redshift, are all above the proton's length for photo pair
    production ( see table \ref{tab:energiaalta}).This requires that the real
    energies of events ( also neglecting the photo-pion production) are shifted towards higher energies:
    consequently, the UHE proton will cross a greater distance and
    will produce more $e^+e^-$ pairs. These pairs will produce
    a so great number of TeV photons that could  exceed the present
    experimental limit on TeV energies. Moreover, the last two
    events in table \ref{tab:energiaalta}, after this shift, have an energy bigger of
    that required for the Z-burst model ($E_{Z-burst}=10^{22}$ eV).

    \item The charged assignments of the first two events for which the correlation
     occurred is zero or negative.

     So a proton can't explain it.

    \item In this model we have mainly a TeV gamma-ray afterglow
    halo surrounding the UHECR event
    because  the  life time  of an electron with an energy
    $E_e=10^{14}eV $ (that produces TeV photons on BBR by ICS) is

\begin{equation}\label{eqn:tempo}
\tau_e \simeq \frac{2.8\times10^{12}}{\gamma}yr\sim
1.5\times10^{4}\biggl(\frac{E}{10^{14}eV}\biggr)^{-1}yr
\end{equation}
which, compared with the proton's time of delay given by eq.
\ref{eqn:ritardo}, is even longer. The electron pair tail in the
direct as well Z-Burst model is very narrow bounded along the
UHECR propagation. Therefore the gamma ring-halo afterglow is well
collimated along UHECR path.
\end{enumerate}
\begin{table*}[htb]
\caption{BL Lacs and their real energies}\label{tab:energiaalta}
\newcommand{\m}{\hphantom{$-$}}
\newcommand{\cc}[1]{\multicolumn{1}{c}{#1}}
\renewcommand{\tabcolsep}{1pc} 
\renewcommand{\arraystretch}{1.2} 
\begin{tabular}{@{}llllll}
\hline
    EGRET Name &  z & d (Mpc)&  $E_{obs} (10^{19}eV)$ & $E_{in} (10^{19}eV)$& Charge assignment \\
    \hline
    0808+5114 & 0.138 & 455 & 3.4 & 9.2 & 0 \\
    1052+5718 & 0.144 & 475 & 7.76 & 14.7 & 0,-1 \\
    1424+3734 & 0.564 & 1861 & 4.97 & $6\times10^{3}$ & 0,+1 \\
    1850+5903 & 0.53 & 1750 & 5.8 &$10^{4}$  & +1 \\
    \hline
\end{tabular}
\end{table*}
 The Z-burst's advantages on the contrary are:
\begin{enumerate}
    \item explanation of the correlation between UHECR and BL
    Lacs, without any charge problems (UHECR maybe either proton or neutron or their anti-particles);
    \item production of gamma-rays precursors and afterglows tails around  GeV
    energy where there aren't strong experimental bounds but
    even some positive evidence shown in correlation with EGRET
    GeV    identified sources,
    \item UHECR calibration with the observed light neutrino
    masses ($0.4-0.1$ eV), well within known atmospheric mass splitting,
    naturally located in an extended hot dark halo within GZK
    distance,
    \item  time delay arrival between the UHECR cluster  compatible with
    observation and with the expected extra-galactic magnetic fields.
\end{enumerate}
Therefore, in conclusion, the Z-burst model is offering pion
decays and gamma GeV signature by synchrotron radiation whose
presence is already found in EGRET; any direct flights of UHECR
from distant BL Lac are often in disagreement with the charge sign
observed and they call even for very high \cite{Neronov2004} TeV
flux halo ; we have shown in present article that TeV signals,
whose fluence might be comparable or larger than UHECR fluxes, are
absent in present data. Therefore in our opinion, the Z-Burst is
at present the most realistic model to explain UHECR events: we
believe that a possible prompt observation of gamma-rays
afterglows surrounding UHECR at MeV-GeV or at TeV energies will
test these two models. We suggest therefore a fast search by
MAGIC, HESS and VERITAS telescopes within a short time scales
(minutes-hours-days or weeks) toward the UHECR arrival directions,
looking for GeVs-TeVs showering halo (electromagnetic imprint of
Z-decay in flight and its synchrotron emission) surrounding the
UHECR event. Also X-Ray search by high resolution X-ray satellites
as  Chandra  pointing promptly toward   UHECR arrivals directions
might find exciting signatures of the shower afterglow along the
particle path.

\end{document}